\documentclass[aps,floatfix,prl,superscriptaddress,reprint,showpacs,10pt]{revtex4-1}
\usepackage[utf8]{inputenc}
\usepackage[dvips]{graphicx}
\usepackage{float}
\usepackage{amssymb}
\usepackage{amsmath}  
\usepackage{dsfont}
\usepackage{color}
\usepackage{array}
\usepackage{subfigure}
\usepackage[locale=US]{siunitx}
\usepackage{bm}
\usepackage{mathrsfs}
\usepackage{pifont}
\usepackage[dvips,pdftitle={Multi-flavor Schwinger model with MPS},pdfauthor={Stefan Kuehn},bookmarks,linkcolor=black,citecolor=black,menucolor=black,urlcolor=black,hyperfootnotes=false,plainpages=false,pdfpagelabels,hypertexnames=false]{hyperref}

\newcommand{\X}{\ding{53}}
\newcommand{\sz}{\ensuremath{\sigma^z}}
\renewcommand{\sp}{\ensuremath{\sigma^+}}
\newcommand{\sm}{\ensuremath{\sigma^-}}

\begin{document}
\title{Density Induced Phase Transitions in the Schwinger Model: A Study with Matrix Product States}
\author{Mari Carmen Ba\~nuls}
\affiliation{Max-Planck-Institut f\"ur Quantenoptik, Hans-Kopfermann-Straße 1, 85748 Garching, Germany}
\author{Krzysztof Cichy}
\affiliation{Goethe-Universit\"at Frankfurt am Main, Institut f\"ur Theoretische Physik,
Max-von-Laue-Stra{\ss}e 1, 60438 Frankfurt am Main, Germany}
\affiliation{Faculty of Physics, Adam Mickiewicz University, Umultowska 85, 61-614 Pozna\'{n}, Poland}
\author{J. Ignacio Cirac}
\affiliation{Max-Planck-Institut f\"ur Quantenoptik, Hans-Kopfermann-Straße 1, 85748 Garching, Germany}
\author{Karl Jansen}
\affiliation{NIC, DESY Zeuthen, Platanenallee 6, 15738 Zeuthen, Germany}
\author{Stefan K\"uhn}
\affiliation{Max-Planck-Institut f\"ur Quantenoptik, Hans-Kopfermann-Straße 1, 85748 Garching, Germany}
\date{\today}
\begin{abstract}
We numerically study the zero temperature phase structure of the multiflavor Schwinger model at nonzero chemical potential. Using matrix product states, we reproduce analytical results for the phase structure for two flavors in the massless case and extend the computation to the massive case, where no analytical predictions are available. Our calculations allow us to locate phase transitions in the mass-chemical potential plane with great precision and provide a concrete example of tensor networks overcoming the sign problem in a lattice gauge theory calculation.
\end{abstract}
\maketitle

Gauge theories are a fundamental concept in high energy physics. Nevertheless, in many cases, such as quantum chromodynamics (QCD), they are notoriously hard, and a full analytical solution seems to be impossible. Following the pioneering work by Wilson~\cite{Wilson1974}, lattice gauge theory (LGT) has become a standard tool for attacking gauge theories in the nonperturbative regime. This discretized formulation on a Euclidean space-time lattice enabled powerful Monte Carlo (MC) simulations that allowed the determination of phase diagrams, mass spectra, and other properties. However, the sign problem~\cite{Troyer2005} prevents accessing certain parameter regimes with this technique, as, for example, large parts of the phase diagram for QCD with chemical potential. Moreover, real-time dynamics are mostly inaccessible, despite some recent progress enabling their study in particular regimes~\cite{Hebenstreit2013}. Consequently, there is an enduring search for alternative approaches overcoming these limitations~\cite{Cristoforetti2013,Gattringer2016,Ammon2016}, among them MC simulations on Lefshetz thimbles, complex Langevin methods, and density of states methods. A different line of research, analyzed in a number of works~\cite{Banerjee2012,Wiese2013,Kuehn2014,Zohar2015a,Dalmonte2016} and recently experimentally realized for small systems~\cite{Martinez2016}, is quantum simulation of gauge theories.

In the last decade, new methods based on tensor networks (TN) have revealed themselves as powerful approaches for the nonperturbative study of quantum many-body systems (see Ref.~\cite{Verstraete2008} for a review), both bosonic and fermionic, without suffering from a sign problem. In the context of LGT, they can be used to approximate the partition function in a Lagrangian formulation~\cite{Takeda2015,Denbleyker2014,Shimizu2014}, but their main power can be exploited in the Hamiltonian formulation, thanks to their capability to efficiently describe the relevant states of the theory~\cite{Banuls2013,Buyens2013,Banuls2015,Banuls2016,Buyens2016,Silvi2016,Kuehn2015,Pichler2015}. Lately, there has been significant theoretical progress with the development of gauge invariant TN formulations suitable for LGT~\cite{Rico2013,Silvi2014,Zohar2015b,Tagliacozzo2014,Haegeman2015,Dalmonte2016}, as well as numerical simulations showing the power of the method for spectral calculations~\cite{Banuls2013,Buyens2013,Buyens2015}, thermal states~\cite{Banuls2015,Banuls2016,Buyens2016}, exploring phase diagrams~\cite{Silvi2016,Zohar2015c}, and simulating real-time evolution for Abelian as well as non-Abelian theories~\cite{Buyens2013,Kuehn2015,Pichler2015}.

Some of these works achieved precisions beyond the reach of MC calculations for the considered models in one spatial dimension. Extending this success to higher spatial dimensions, although conceptually possible, is not an immediate task in the general case, but in regimes where MC simulations suffer from the sign problem, TN techniques should provide a very general solution. This major promise can already be demonstrated in the one-dimensional case, a task that we tackle in this Letter. We study the multiflavor Schwinger model (quantum electrodynamics in 1+1 dimension) at nonzero chemical potential and perform calculations in regimes where MC calculations would suffer from a sign problem \footnote{It has been noticed that in certain restricted parameter regimes, the sign problem can be circumvented~\cite{Gattringer2015}, but here we adopt a general prescription, common for massless and massive cases, where that is not the case.}. We go through the full extrapolation procedure to recover the continuum limit to explicitly show the power of TN approaches for overcoming the sign problem.

For two flavors with equal masses, the case on which we focus here, the model has an SU(2) isospin symmetry between the flavors and is in many aspects similar to QCD as it shows confinement, an anomalous U(1) current in the massless limit and a nonvanishing chiral condensate. In Refs.~\cite{Narayanan2012,Lohmayer2013}, it was found analytically that at zero temperature the model supports an infinite number of phases characterized by the isospin number and separated by first-order phase transitions.   

Here, we numerically study the Hamiltonian lattice formulation of the model with matrix product states (MPS) and extrapolate to the continuum limit. As a first necessary step, we reproduce the analytical prediction for massless fermions from Refs.~\cite{Narayanan2012,Lohmayer2013} with great precision. Furthermore, our calculation can be readily extended to the massive case, where no analytical computations are available, and we observe that the phase structure changes significantly. Using the MPS approach, and considering the case of vanishing background field, we are able to map out accurately the phase diagram of the model in the mass-chemical potential plane for a fixed volume. Our results thus constitute an explicit demonstration that MPS allow reliable numerical simulations in a regime where the MC approach would suffer from the sign problem.

We adopt a lattice formulation with Kogut-Susskind staggered fermions~\cite{Kogut1975}. In the temporal gauge, and in absence of a background field, the Hamiltonian for $F$ flavors on a lattice with spacing $a$ and $N$ sites reads
\begin{align}
 \begin{aligned}
 H = &-\frac{i}{2a}\sum_{n=0}^{N-2}\sum_{f=0}^{F-1}\left(\phi^\dagger_{n,f}e^{i\theta_n}\phi_{n+1,f}-\mathrm{h.c.}\right)\\
 &+\sum_{n=0}^{N-1}\sum_{f=0}^{F-1}\left(m_f(-1)^n +\kappa_f \right)\phi^\dagger_{n,f}\phi_{n,f}\\
 &+ \frac{ag^2}{2}\sum_{n=0}^{N-2} L_n^2.
 \end{aligned}
 \label{hamiltonian}
\end{align}
Here, $\phi_{n,f}$ is a single component fermionic field describing a fermion of flavor $f$ on site $n$, and $m_f/g$ and $\kappa_f/g$ are the corresponding mass and chemical potential in units of the coupling constant, $g$. The operators $L_n$ and $\theta_n$ act on the gauge links between the fermions and $L_n$ gives the electric flux on link $n$. They are canonical conjugates, $[\theta_n,L_m]=i\delta_{n,m}$; hence, $e^{i\theta_n}$ acts as a rising operator for the electric flux. We work with a compact formulation, where $\theta_n$ is restricted to $[0,2\pi]$~\cite{Hamer1997}. 

Physical states, $|\psi\rangle$, have to satisfy the Gauss law, $G_n|\psi\rangle=0$ $\forall n$, where $G_n = L_n - L_{n-1} - \sum_{f=0}^{F-1}\left( \phi^\dagger_{n,f}\phi_{n,f}-\frac{1}{2}(1-(-1)^n)\right)$ are the generators for gauge transformations. For open boundary conditions (OBC), this allows us to integrate out the gauge fields. Assuming zero electric field on the left boundary, applying a residual gauge transformation and with a rescaling that makes it dimensionless~\cite{Banks1976}, the Hamiltonian (\ref{hamiltonian}) can be written as
\begin{align}
\begin{aligned}
 W =& -ix\sum_{n=0}^{N-2}\sum_{f=0}^{F-1}\left(\phi^\dagger_{n,f}\phi_{n+1,f}-\mathrm{h.c.}\right)\\
 &+\sum_{n=0}^{N-1}\sum_{f=0}^{F-1}\left(\mu_f(-1)^n +\nu_f \right)\phi^\dagger_{n,f}\phi_{n,f} \\
 &+ \sum_{n=0}^{N-2} \left( \sum_{k=0}^n\left(\sum_{f=0}^{F-1}\phi_{k,f}^\dagger\phi_{k,f}-\frac{F}{2}(1-(-1)^k)\right)\right)^2,
\end{aligned}
\label{hamiltonian_dimensionless}
\end{align}
where the adimensional parameters of the problem are $x=1/(ag)^2$, $\mu_f = 2\sqrt{x}m_f/g$, and $\nu_f = 2\sqrt{x}\kappa_f/g$. In the following, we will focus on the case of two flavors in the sector of vanishing total charge, for which the conventional MC approach in general suffers from the sign problem~\footnote{In the special case $\nu_0+\nu_1=0$, the sign problem can be circumvented~\cite{Narayanan2012}.}. 

Our variational ansatz is a MPS with OBC. For $N$ sites this is a state of the form
\begin{align*}
 |\psi\rangle = \sum_{i_0,i_1,\dots i_{N-1}} A^{i_0}_0A^{i_1}_1 \dots A^{i_{N-1}}_{N-1}|i_0\rangle \otimes \dots\otimes |i_{N-1}\rangle,
\end{align*}
where $|i_k\rangle_{i_k=1}^d$ is a basis for the Hilbert space on site $k$, $A^{i_k}_k$  are complex $D\times D$ matrices for $0<k<N-1$, and $A^{i_0}_0$ ($A^{i_{N-1}}_{N-1}$) is a $D$-dimensional row (column) vector. The bond dimension of the MPS, $D$, determines the number of variational parameters and limits the maximum entanglement in the state (see, e.g., Ref.~\cite{Verstraete2008}).

Although Hamiltonian (\ref{hamiltonian_dimensionless}) is nonlocal, it can be expected that MPS are good ans\"atze for the ground state, as the original model is local, and its low-energy states are characterized by small electric field values~\footnote{As the electric field for low energy states is small, the gauge links can be effectively considered as finite dimensional. Hence, as the Hamiltonian (\ref{hamiltonian}) is local, the arguments from Ref.~\cite{Verstraete2006} apply, showing that even for the critical case, its ground state can be well approximated by a MPS with a small bond dimension. Integrating out the gauge field only moderately increases the bond dimension (more specifically, projecting the ground state of Hamiltonian (\ref{hamiltonian}) to the physical subspace increases it at maximum by a factor on the order of the effective dimension of the links). Therefore, also the ground state of Hamiltonian (\ref{hamiltonian_dimensionless}) is expected to be well described by a MPS with small bond dimension.}. To show that MPS allow for reliable calculations with proper continuum limit in the regime of the sign problem, we first reproduce the analytical predictions for the massless case from Refs.~\cite{Narayanan2012,Lohmayer2013}, which studied the continuum model in a fixed volume. Consequently, we consider lattices of constant volume, $Lg=N/\sqrt{x}$. The isospin number on the lattice is given by $\Delta N = N_0 - N_1$, with $N_i = \sum_{n=0}^{N-1}\phi_{n,i}^\dagger\phi_{n,i}$. It can be shown that the Hamiltonian (\ref{hamiltonian_dimensionless}) up to a constant only depends on the difference  $\nu_1-\nu_0$, commonly called the isospin chemical potential in the literature (see Supplemental Material). Thus, we study $\Delta N$ in the ground state as a function of the difference between the chemical potentials. Following Refs.~\cite{Narayanan2012,Lohmayer2013}, we define the rescaled isospin chemical potential $\mu_I/2\pi = N(\nu_1-\nu_0)/4\pi x$, and hereafter, we fix $\nu_0=0$ and only vary $\nu_1$. We are thus studying the model in a situation where the MC approach suffers from the sign problem. To probe for possible finite volume effects, we explore $Lg = 2,6,8$. 

In order to be able to extrapolate to the continuum limit, we study several lattice spacings corresponding to $x\in[9,121]$. MPS calculations are subject to a truncation error due to the limited bond dimension reachable, bounded by the computational cost of treating too large matrices in the ansatz. To control this error for each combination of $(Lg,x,\mu_I/2\pi)$, we repeat the computation for several bond dimensions, $D\in[40,220]$ and extrapolate to $D\to \infty$ (see  Supplemental Material). Although MPS and TN in general can describe fermionic degrees of freedom, we map Eq. (\ref{hamiltonian_dimensionless}) to a spin chain by a Jordan-Wigner transformation for convenience in the numerical simulations (see Supplemental Material).

\begin{figure}[htp!]
\centering
\includegraphics[width=0.49\textwidth]{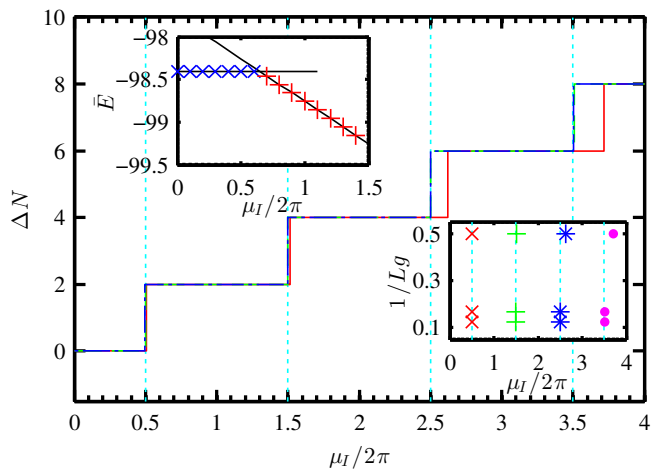}
\caption{Continuum estimate for $\Delta N$ versus $\mu_I/2\pi$, for volumes 2 (red solid), 6 (green dashed), and 8 (blue dash-dotted line). The vertical lines indicate the theoretical prediction for the phase transitions in the massless case. Upper inset: Close-up around the first transition for $Lg=8$, $x=16$, $m/g=0$, $D=160$. Shown are MPS results for $\Delta N=0$ (blue crosses), $\Delta N=2$ (red \X's), and the corresponding predictions (solid lines). Lower inset: Volume dependence of the continuous location of the transitions for the first (red \X's), second (green crosses), third (blue asterisks), and fourth (magenta dots) transition.}
\label{fig:jumplocsBond_mg0}
\end{figure}

The results for the massless case are shown in Fig. \ref{fig:jumplocsBond_mg0}. As $\mu_I/2\pi$ is increased, $\Delta N$ exhibits discontinuous changes, corresponding to the crossing of the lowest energy levels for two different isospin numbers. This leads to an abrupt  change of the nature of the ground state, indicated by first-order (discontinuous) quantum phase transitions between phases characterized by their isospin number. The location of the transition is determined by the position of the energy cusps on the $\mu_I/2\pi$ axis, as seen in the upper inset of Fig. \ref{fig:jumplocsBond_mg0}. Repeating the calculations for several lattice spacings, we can estimate the continuum phase structure of the model (see Supplemental Material). For the first two transitions, our results do not show any volume dependence, in agreement with Refs.~\cite{Narayanan2012,Lohmayer2013}. However, for transitions between phases with larger $\Delta N$, we can see that for $Lg=2$, there are deviations due to finite volume effects. For $Lg\geq6$, those disappear, and we recover the analytical results in the entire parameter regime under study. We conclude that the transitions occur for $\mu_I/2\pi$ values which are odd multiples of $1/2$, in agreement with the analytical results. The finite volume effects found in our MPS calculation for small $Lg$ can be explained because the total fermion number coincides with the number of sites, $N_0+N_1=N$. Hence, the system size ultimately upper bounds $N_i$, and larger values for $\Delta N$ at a fixed volume would require larger system sizes and correspondingly, larger values of $x$ to reach the correct continuum limit.

\begin{figure}[htp!]
\centering
\includegraphics[width=0.49\textwidth]{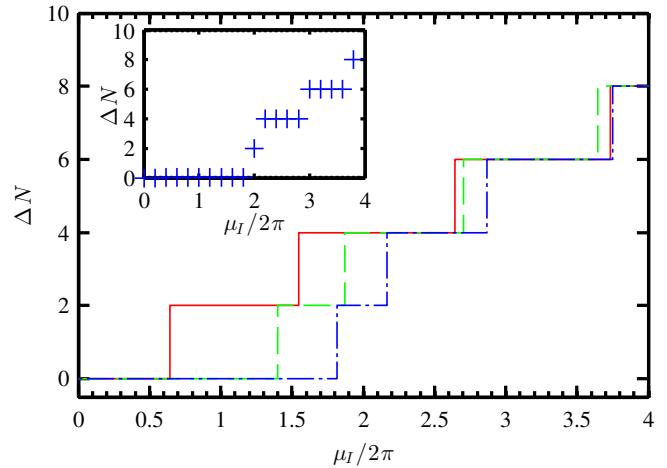}
\caption{Continuum estimate for $\Delta N$ versus $\mu_I/2\pi$, for volumes 2 (red solid), 6 (green dashed), and 8 (blue dash-dotted line). Inset: Isospin number versus $\mu_I/2\pi$ for $Lg=8$, $x=121$, $m/g=0.5$, $D=220$.}
\label{fig:jumplocsBond_mg05}
\end{figure}

In contrast to the analytical calculation in Refs.~\cite{Narayanan2012,Lohmayer2013}, the MPS formalism can deal with (arbitrary) mass values. Proceeding in the same way for $m/g=0.5$, we obtain the results shown in Fig. \ref{fig:jumplocsBond_mg05}. We observe that the new energy scale introduced by $m/g$ leads to a change in the phase structure, as the transitions are not equidistantly spaced anymore. The continuum estimates show a clear volume dependence, even for the first transition, and the size of the plateaus is no longer fixed.

\begin{figure}[htp!]
\centering
\includegraphics[width=0.49\textwidth]{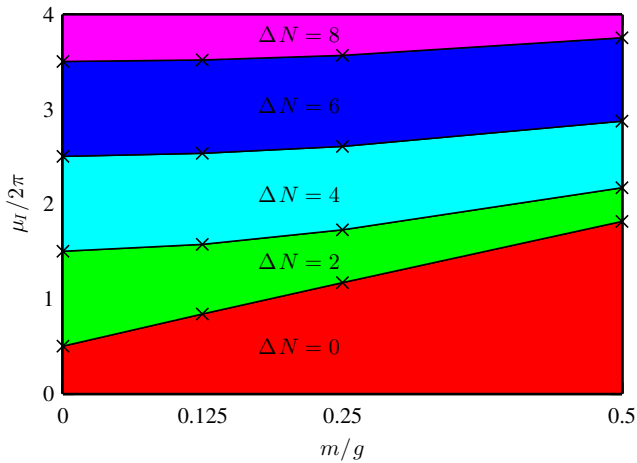}
\caption{Phase diagram in the $m/g$ - $\mu_I/2\pi$ plane for $Lg=8$. The black \X's mark the computed data points, the different colors indicate the different phases.}
\label{fig:phasediagram}
\end{figure}

Computing the phase structure for several masses, we can map out the phase diagram for the model in the $m/g$ - $\mu_I/2\pi$ plane for a fixed volume. Figure  \ref{fig:phasediagram} shows the results for $Lg=8$. For larger masses, the phase characterized by $\Delta N=0$ survives up to larger values of $\mu_I/2\pi$, and the size of the region for the $\Delta N= 2$ phase shrinks. The regions describing phases with larger $\Delta N$ are less affected and only slightly bend towards higher values of the chemical potential difference. This behavior can be understood qualitatively as follows: the energy eigenvalues inside each phase only depend on the chemical potential difference, up to a constant (see Fig.~\ref{fig:jumplocsBond_mg0}). This constant is mass dependent, and comparing its value at nonzero $m/g$ to the massless case, we observe larger changes for phases characterized by a small isospin number. Consequently, the locations of the level crossings, and hence the locations of the phase transitions, are shifted, especially for phases characterized by small $\Delta N$ (see Supplemental Material).

\begin{figure}[htp!]
  \centering
  \includegraphics[width=0.49\textwidth]{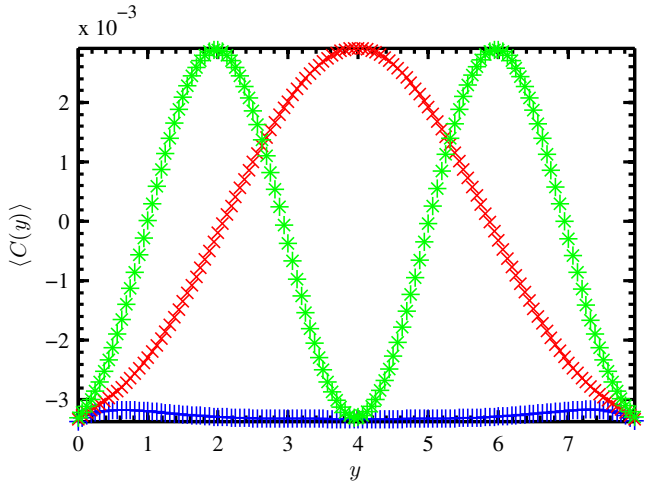}
\caption{$\langle C(y)\rangle$ for $Lg=8$, $x=1024$, $m/g=0$, $D=160$, and different phases. The blue crosses represent $\Delta N=0$, the red \X's $\Delta N=2$, and the green asterisks $\Delta N=4$.}
\label{fig:c_oscillations}
\end{figure}

The MPS method is not only free from the sign problem, but, at the end of the computation, it also yields the ground state wave function, hence giving easy access to observables that can be expressed as matrix product operators~\cite{McCulloch2007}. An interesting observable is the chiral condensate. Previous studies~\cite{Fischler1979,Kao1994,Christiansen1996} for the (single-flavor) Schwinger model found that at finite density, the chiral condensate shows spatial inhomogeneities of the form $\langle \bar{\psi}(y)\psi(y)\rangle = \langle\bar{\psi}\psi\rangle_0\cos(2\kappa y)$, where $\psi$ is a two component Dirac spinor, $\kappa$ is the chemical potential, $y$ the position, and $\langle\bar{\psi}\psi\rangle_0$ the (spatially homogeneous) expectation value of the chiral condensate for vanishing chemical potential. Later work instead argued that these oscillations occur due to the breaking of translational invariance in finite systems~\cite{Metlitski2007}. To be able to compare our staggered lattice calculation to these continuum results, we sum the contribution of an even and its neighboring odd site to the chiral condensate and look at $C(y=2n/\sqrt{x})=\sum_{f=0}^{F-1}(C_{n,f} + C_{n+1,f})$, $n$ even, where $C_{n,f} = \frac{\sqrt{x}}{N} (-1)^n\phi^\dagger_{n,f}\phi_{n,f}$~\footnote{The oscillations are also present for each individual flavor, nevertheless, for convenience in the visualization we sum both flavors.}. The result for $Lg=8$ in the massless case is shown in Fig. \ref{fig:c_oscillations}.
The value at zero density (corresponding to the $\Delta N=0$ phase) is homogeneous up to small finite size effects at the boundaries, as expected from the theoretical result. For phases at nonzero density (given by $\Delta N\neq 0$), the condensate starts to oscillate sinusoidally, as expected for a finite system breaking translational invariance, and we observe an increase in the oscillation frequency with increasing density. The oscillation amplitudes are close to $\langle C(y)\rangle_0$, similar to the theoretical predictions from Refs.~\cite{Fischler1979,Kao1994,Christiansen1996,Metlitski2007} for the single-flavor case. A more detailed study of the oscillations in the chiral condensate will be shown elsewhere~\cite{Banuls2016b}.

In summary, we have shown a successful lattice calculation in the regime where the conventional MC approach suffers from the sign problem. Our results for the massless case in a sufficiently large volume agree with great precision with the analytical calculations from Refs.~\cite{Narayanan2012,Lohmayer2013}, and we recover the predicted phase structure and locations of the phase transitions after extrapolating to the continuum limit. Furthermore, our calculations can be immediately extended to the massive case, where no analytical results are available. In this case, the observed phase structure is significantly different, and the locations of the phase transitions are no longer independent of $Lg$. We can map out the phase diagram of the model at a fixed volume in the $m/g$ - $\mu_I/2\pi$ plane, and we see that the transition from $\Delta N=0$ to $\Delta N=2$ is significantly shifted towards higher values of the chemical potential at the expense of the phase characterized by $\Delta N=2$. Phases with larger values of $\Delta N$ are less affected and only slightly shifted towards higher values of $\mu_I/2\pi$ for increasing mass. Our results for the condensate are very similar to the theoretical predictions for the single-flavor case at nonzero density. We observe oscillations with a density dependent frequency around zero with an amplitude close to the zero density condensate value.

In our study, we focused on the phases at zero background field and temperature, with nonvanishing chemical potential, to explore a regime that suffers from the sign problem in conventional MC calculations. Notice, however, that the model also exhibits interesting features in other parameter regimes. In particular, in the absence of chemical potential and background field, it has been shown to have a second-order phase transition for zero fermion mass at $T_c=0$ ~\cite{Smilga1996,Duerr2000}. It might also show a transition, similar to the single-flavor case, at a nonvanishing background field, as has been argued in Ref.~\cite{Hosotani1998}. Adding a background field as well as a generalization to a nonzero temperature~\cite{Banuls2015,Banuls2016,Buyens2016} is straightforward; hence, these regimes are also amenable to TN studies \footnote{Notice that MPS are not limited to the finite volume case studied here, but also allow us to access observables in the thermodynamic limit, as has been demonstrated in Refs.~\cite{Banuls2013,Buyens2013,Buyens2015,Buyens2016,Banuls2015,Banuls2016}.}.

The MPS approach can be easily extended to an arbitrary number of flavors (see Supplemental Material). To some extent, it is also possible to simulate real-time evolution~\cite{Buyens2013} and thus, to address dynamical aspects of the model. Additionally, our results can serve as a test bench for other methods trying to overcome the sign problem. Moreover, our study is also promising for higher dimensions. For the same reasons MPS with small bond dimension provide a good ansatz for the one-dimensional case, we expect that the low-energy states for the two-dimensional case can be efficiently described by projected entangled pair states (PEPS)~\cite{Verstraete2004b}, the generalization of MPS to two dimensions. The remarkable progress in the analytical~\cite{Zohar2015c,Zohar2016} and numerical techniques~\cite{Phien2015,Corboz2016a,Vanderstraeten2016a,Liu2016} for PEPS is bringing this closer to realization.

\begin{acknowledgments}
K.C.\ was supported by the Deutsche Forschungsgemeinschaft (DFG), Project No. CI 236/1-1 (Sachbeihilfe).
\end{acknowledgments}
\bibliographystyle{h-physrev}
\bibliography{Papers_MPQ}

\begin{thebibliography}{10}

\bibitem{Wilson1974}
K.~G. Wilson,
\newblock Phys. Rev. D {\bf 10}, 2445 (1974).

\bibitem{Troyer2005}
M.~Troyer and U.-J. Wiese,
\newblock Phys. Rev. Lett. {\bf 94}, 170201 (2005).

\bibitem{Hebenstreit2013}
F.~Hebenstreit, J.~Berges, and D.~Gelfand,
\newblock Phys. Rev. Lett. {\bf 111}, 201601 (2013).

\bibitem{Cristoforetti2013}
M.~Cristoforetti, F.~Di~Renzo, A.~Mukherjee, and L.~Scorzato,
\newblock Phys. Rev. D {\bf 88}, 051501 (2013).

\bibitem{Gattringer2016}
C.~Gattringer and K.~Langfeld,
\newblock Int. J. Mod. Phys. A {\bf 31}, 1643007 (2016).

\bibitem{Ammon2016}
A.~Ammon, T.~Hartung, K.~Jansen, H.~Le\"ovey, and J.~Volmer,
\newblock Phys. Rev. D {\bf 94}, 114508 (2016).

\bibitem{Banerjee2012}
D.~Banerjee {\em et~al.},
\newblock Phys. Rev. Lett. {\bf 109}, 175302 (2012).

\bibitem{Wiese2013}
U.-J. Wiese,
\newblock Ann. Phys. (Amsterdam) {\bf 525}, 777 (2013).

\bibitem{Kuehn2014}
S.~K\"uhn, J.~I. Cirac, and M.-C. Ba\~nuls,
\newblock Phys. Rev. A {\bf 90}, 042305 (2014).

\bibitem{Zohar2015a}
E.~Zohar, J.~I. Cirac, and B.~Reznik,
\newblock Rep. Prog. Phys. {\bf 79}, 014401 (2016).

\bibitem{Dalmonte2016}
M.~Dalmonte and S.~Montangero,
\newblock Contemp. Phys. {\bf 57}, 388 (2016).

\bibitem{Martinez2016}
E.~A. Martinez {\em et~al.},
\newblock Nature {\bf 534}, 516–519 (2016).

\bibitem{Verstraete2008}
F.~Verstraete, V.~Murg, and J.~Cirac,
\newblock Adv. Phys. {\bf 57}, 143 (2008).

\bibitem{Takeda2015}
S.~Takeda and Y.~Yoshimura,
\newblock Prog. Theor. Exp. Phys. {\bf 2015} (2015).

\bibitem{Denbleyker2014}
A.~Denbleyker {\em et~al.},
\newblock Phys. Rev. D {\bf 89}, 016008 (2014).

\bibitem{Shimizu2014}
Y.~Shimizu and Y.~Kuramashi,
\newblock Phys. Rev. D {\bf 90}, 014508 (2014).

\bibitem{Banuls2013}
M.~C. Ba\~{n}uls, K.~Cichy, K.~Jansen, and J.~I. Cirac,
\newblock J. High Energy Phys. {\bf 2013}, 158 (2013).

\bibitem{Buyens2013}
B.~Buyens, J.~Haegeman, K.~Van~Acoleyen, H.~Verschelde, and F.~Verstraete,
\newblock Phys. Rev. Lett. {\bf 113}, 091601 (2014).

\bibitem{Banuls2015}
M.~C. Ba\~{n}uls, K.~Cichy, J.~I. Cirac, K.~Jansen, and H.~Saito,
\newblock Phys. Rev. D {\bf 92}, 034519 (2015).

\bibitem{Banuls2016}
M.~C. Ba\~{n}uls, K.~Cichy, K.~Jansen, and H.~Saito,
\newblock Phys. Rev. D {\bf 93}, 094512 (2016).

\bibitem{Buyens2016}
B.~Buyens, F.~Verstraete, and K.~Van~Acoleyen,
\newblock Phys. Rev. D {\bf 94}, 085018 (2016).

\bibitem{Silvi2016}
P.~Silvi, E.~Rico, M.~Dalmonte, F.~Tschirsich, and S.~Montangero,
\newblock arXiv:1606.05510  (2016).

\bibitem{Kuehn2015}
S.~K\"uhn, E.~Zohar, J.~Cirac, and M.~C. Ba\~{n}uls,
\newblock J. High Energy Phys. {\bf 2015}, 130 (2015).

\bibitem{Pichler2015}
T.~Pichler, M.~Dalmonte, E.~Rico, P.~Zoller, and S.~Montangero,
\newblock Phys. Rev. X {\bf 6}, 011023 (2016).

\bibitem{Rico2013}
E.~Rico, T.~Pichler, M.~Dalmonte, P.~Zoller, and S.~Montangero,
\newblock Phys. Rev. Lett. {\bf 112}, 201601 (2014).

\bibitem{Silvi2014}
P.~Silvi, E.~Rico, T.~Calarco, and S.~Montangero,
\newblock New J. Phys. {\bf 16}, 103015 (2014).

\bibitem{Zohar2015b}
E.~Zohar and M.~Burrello,
\newblock New J. Phys. {\bf 18}, 043008 (2016).

\bibitem{Tagliacozzo2014}
L.~Tagliacozzo, A.~Celi, and M.~Lewenstein,
\newblock Phys. Rev. X {\bf 4}, 041024 (2014).

\bibitem{Haegeman2015}
J.~Haegeman, K.~Van~Acoleyen, N.~Schuch, J.~I. Cirac, and F.~Verstraete,
\newblock Phys. Rev. X {\bf 5}, 011024 (2015).

\bibitem{Buyens2015}
B.~Buyens, J.~Haegeman, H.~Verschelde, F.~Verstraete, and K.~Van~Acoleyen,
\newblock Phys. Rev. X {\bf 6}, 041040 (2016).

\bibitem{Zohar2015c}
E.~Zohar, M.~Burrello, T.~B. Wahl, and J.~I. Cirac,
\newblock Ann. Phys. (Amsterdam) {\bf 363}, 385  (2015).

\bibitem{Note1}
It has been noticed that in certain restricted parameter regimes, the sign
  problem can be circumvented~\cite {Gattringer2015}, but here we adopt a
  general prescription, common for massless and massive cases, where that is
  not the case.

\bibitem{Narayanan2012}
R.~Narayanan,
\newblock Phys. Rev. D {\bf 86}, 125008 (2012).

\bibitem{Lohmayer2013}
R.~Lohmayer and R.~Narayanan,
\newblock Phys. Rev. D {\bf 88}, 105030 (2013).

\bibitem{Kogut1975}
J.~Kogut and L.~Susskind,
\newblock Phys. Rev. D {\bf 11}, 395 (1975).

\bibitem{Hamer1997}
C.~J. Hamer, Z.~Weihong, and J.~Oitmaa,
\newblock Phys. Rev. D {\bf 56}, 55 (1997).

\bibitem{Banks1976}
T.~Banks, L.~Susskind, and J.~Kogut,
\newblock Phys. Rev. D {\bf 13}, 1043 (1976).

\bibitem{Note2}
In the special case $\nu _0+\nu _1=0$, the sign problem can be
  circumvented~\cite {Narayanan2012}.

\bibitem{Note3}
As the electric field for low energy states is small, the gauge links can be
  effectively considered as finite dimensional. Hence, as the Hamiltonian (\ref
  {hamiltonian}) is local, the arguments from Ref.~\cite {Verstraete2006}
  apply, showing that even for the critical case, its ground state can be well
  approximated by a MPS with a small bond dimension. Integrating out the gauge
  field only moderately increases the bond dimension (more specifically,
  projecting the ground state of Hamiltonian (\ref {hamiltonian}) to the
  physical subspace increases it at maximum by a factor on the order of the
  effective dimension of the links). Therefore, also the ground state of
  Hamiltonian (\ref {hamiltonian_dimensionless}) is expected to be well
  described by a MPS with small bond dimension.

\bibitem{McCulloch2007}
I.~P. McCulloch,
\newblock J. Stat. Mech. {\bf 2007}, P10014 (2007).

\bibitem{Fischler1979}
W.~Fischler, J.~Kogut, and L.~Susskind,
\newblock Phys. Rev. D {\bf 19}, 1188 (1979).

\bibitem{Kao1994}
Y.-C. Kao and Y.-W. Lee,
\newblock Phys. Rev. D {\bf 50}, 1165 (1994).

\bibitem{Christiansen1996}
H.~R. Christiansen and F.~A. Schaposnik,
\newblock Phys. Rev. D {\bf 53}, 3260 (1996).

\bibitem{Metlitski2007}
M.~A. Metlitski,
\newblock Phys. Rev. D {\bf 75}, 045004 (2007).

\bibitem{Note4}
The oscillations are also present for each individual flavor, nevertheless, for
  convenience in the visualization we sum both flavors.

\bibitem{Banuls2016b}
M.~C. Ba\~{n}uls {\em et~al.},
\newblock arXiv:1611.01458; PoS(LATTICE 2016)316  (2016).

\bibitem{Smilga1996}
A.~Smilga and J.~J.~M. Verbaarschot,
\newblock Phys. Rev. D {\bf 54}, 1087 (1996).

\bibitem{Duerr2000}
S.~D\"urr,
\newblock arXiv:hep-th/0009094  (2000).

\bibitem{Hosotani1998}
Y.~Hosotani and R.~Rodriguez,
\newblock J. Phys. A {\bf 31}, 9925 (1998).

\bibitem{Note5}
Notice that MPS are not limited to the finite volume case studied here, but
  also allow us to access observables in the thermodynamic limit, as has been
  demonstrated in Refs.~\cite
  {Banuls2013,Buyens2013,Buyens2015,Buyens2016,Banuls2015,Banuls2016}.

\bibitem{Verstraete2004b}
F.~Verstraete and J.~I. Cirac,
\newblock arXiv:cond-mat/0407066  (2004).

\bibitem{Zohar2016}
E.~Zohar, T.~B. Wahl, M.~Burrello, and J.~I. Cirac,
\newblock Ann. Phys. (Amsterdam) {\bf 374}, 84  (2016).

\bibitem{Phien2015}
H.~N. Phien, J.~A. Bengua, H.~D. Tuan, P.~Corboz, and R.~Or\'us,
\newblock Phys. Rev. B {\bf 92}, 035142 (2015).

\bibitem{Corboz2016a}
P.~Corboz,
\newblock Phys. Rev. B {\bf 94}, 035133 (2016).

\bibitem{Vanderstraeten2016a}
L.~Vanderstraeten, J.~Haegeman, P.~Corboz, and F.~Verstraete,
\newblock Phys. Rev. B {\bf 94}, 155123 (2016).

\bibitem{Liu2016}
W.-Y. Liu, S.-J. Dong, Y.-J. Han, G.-C. Guo, and L.~He,
\newblock arXiv:1611.09467  (2016).

\bibitem{Gattringer2015}
C.~Gattringer, T.~Kloiber, and V.~Sazonov,
\newblock Nucl. Phys. B {\bf 897}, 732  (2015).

\bibitem{Verstraete2006}
F.~Verstraete and J.~I. Cirac,
\newblock Phys. Rev. B {\bf 73}, 094423 (2006).

\end{thebibliography}

\newpage
\widetext
\begin{center}
\textbf{Supplemental Material for ``Density Induced Phase Transitions in the Schwinger Model: A Study with Matrix Product States''}
\end{center}
\setcounter{equation}{0}
\setcounter{figure}{0}
\setcounter{table}{0}
\makeatletter
\renewcommand{\theequation}{S\arabic{equation}}
\renewcommand{\thefigure}{S\arabic{figure}}
\renewcommand{\bibnumfmt}[1]{[S#1]}
\renewcommand{\citenumfont}[1]{S#1}

\section{Spin formulation\label{app:spin_formulation}}
For convenience in the simulations, we use an equivalent spin formulation for the Hamiltonian from Eq. (\ref{hamiltonian_dimensionless}) obtained via a Jordan-Wigner transformation
\begin{align*}
 \phi_k = \prod_{l<k}(i\sz_l)\sm_k,\quad\quad\quad
 \phi_k^\dagger &= \prod_{l<k}(-i\sz_l)\sp_k,
\end{align*}
where we choose to order the fermions inside each site according to their flavor such that $\phi_{n,f} = \phi_{nF+f}$. In the formula above $\sigma_j^z$ and $\sigma_j^\pm$ are the usual Pauli matrices acting on spin $j$. The Hamiltonian in spin language is given by
\begin{align}
\begin{aligned}
 W=&-x\sum_{p=0}^{NF-1}\left(\sp_{p}(i\sz_{p+1})\dots (i\sz_{p+F-1})\sm_{p+F} + \mathrm{h.c.}\right)\\
 &+\sum_{n=0}^{N-1}\sum_{f=0}^{F-1}\bigl(\mu_f(-1)^n +\nu_f \bigr)\frac{1+\sz_{nF+f}}{2}\\
 &+\sum_{n=0}^{N-2}\left(\frac{F}{2}\sum_{k=0}^n(-1)^k + \frac{1}{2}\sum_{k=0}^n\sum_{f=0}^{F-1}\sz_{kF+f}\right)^2,
\end{aligned}
\label{spin_hamiltonian_raw}
\end{align}
hence for a system with $N$ sites and $F$ flavors of fermions, we end up with a spin chain of length $NF$ after the transformation.

Additionally we are interested in the sector with vanishing total charge. To impose that, we add a penalty term $P = \lambda\left(\sum_{n=0}^{N-1} Q_n\right)^2$ to the Hamiltonian from Eq. (\ref{spin_hamiltonian_raw}), where $Q_n$ is the staggered charge given by $Q_n = \sum_{f=0}^{F-1}\frac{1}{2}\left(\sz_{nF+f}+(-1)^n\right)$ in the spin formulation. The Hamiltonian including the penalty term for vanishing total charge can be implemented efficiently as matrix product operator with a bond dimension $D'=2F+3$, despite the long range interactions. 

For our calculations presented in the main text, we chose $\lambda=1000$ and checked the expectation value of $P$, where we found that it is negligible for all our simulations.

\section{Extracting the locations of phase transitions\label{app:jump_locs}}
\begin{figure}[htp!]
  \centering
  \includegraphics[width=0.5\textwidth]{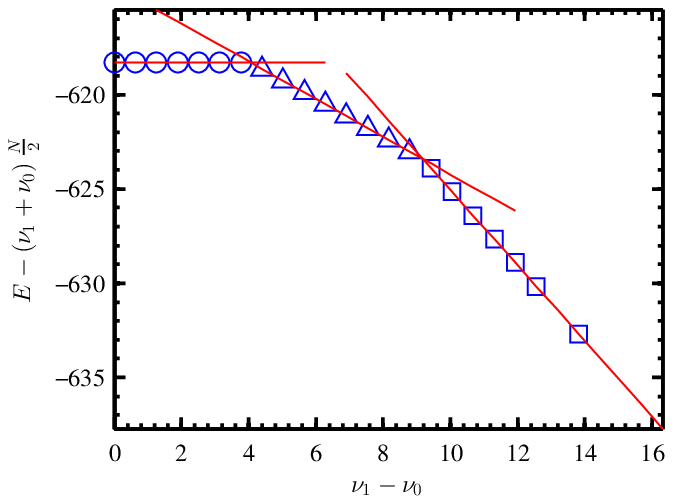}
\caption{Ground state energy as a function of the chemical potential difference for $m/g=0$, $Lg=8$, $x=16$, and $D=160$. The different symbols correspond to $\Delta N=0$ (circles), $\Delta N=2$ (triangles) and $\Delta N=4$ (squares). The lines represent linear functions with slope $p_{(N,\Delta N)}$.}
\label{fig:last_step_mc}
\end{figure}

Here we briefly explain how we extract the locations of the phase transitions for the two-flavor case. A short calculation shows that the Hamiltonian in the sector of vanishing total charge conserves $N_0$ and $N_1$ as well as $N=N_0+N_1$. Hence it is block diagonal and the blocks can be labeled with $(N,\Delta N = N_0-N_1)$. Inside a block the chemical potential terms are proportional to the identity and the Hamiltonian can be written as 
 \begin{align*}
 W =\nu_0N_0 + \nu_1N_1 + W_\mathrm{aux},
\end{align*}
where $W_\mathrm{aux}$ sums up all remaining terms that are independent of the chemical potential. The ground state energy of this Hamiltonian is given by
\begin{align}
 E_{(N,\Delta N)}(\nu_0,\nu_1) &= \nu_0N_0 + \nu_1N_1 + E_\mathrm{min}(W_\mathrm{aux}|_{(N,\Delta N)}) \label{Eblock}   \\
               &= \frac{N}{2}\left(\nu_0+\nu_1\right) - \underbrace{\frac{\Delta N}{2}}_{p_{(N,\Delta N)}}\left(\nu_1-\nu_0\right) + E_\mathrm{min}(W_\mathrm{aux}|_{(N,\Delta N)}). \label{Eblockscaling}
\end{align}
where $E_\mathrm{min}(W_\mathrm{aux}|_{(N,\Delta N)})$ is a block dependent, i.e. isospin number dependent constant. From the equation above, one can immediately see that having a single value for $E_{(N,\Delta N)}(\nu_0,\nu_1)$ available inside each block is enough to determine this constant. Moreover, Eq. (\ref{Eblockscaling}) reveals that for fixed $N$ the energy inside each block only depends linearly on $\nu_1-\nu_0$ up to a (chemical potential dependent) constant, with a slope proportional to $\Delta N$ (see Fig. \ref{fig:last_step_mc}). 

A phase transition, and hence a discontinuity in the isospin number, occurs, if it is energetically favorable to go from one block characterized by $(N,\Delta N)$ to a neighboring block characterized by $(N,\Delta \bar{N}=\Delta N\pm 2)$. As discussed above, inside each block the energy scales linearly (up to a constant) with a block dependent slope. Thus a phase transition corresponds to the intersection point of the two linear functions describing the energy inside these blocks, as can be seen in Fig. \ref{fig:last_step_mc}. 

Equating $E_{(N,\Delta N)}(\nu_0,\nu_1) = E_{(N,\Delta \bar{N})}(\nu_0,\nu_1)$ and using Eq. (\ref{Eblockscaling}) we can obtain the following analytical expression for the intersection points:
\begin{align}
 (\nu_1-\nu_0)|_\mathrm{jump} &= \frac{E_\mathrm{min}(W_\mathrm{aux}|_{(N,\Delta \bar{N})}) - E_\mathrm{min}(W_\mathrm{aux}|_{(N,\Delta N)})}{p_{(N,\Delta \bar{N})} - p_{(N,\Delta N)}}\label{jumplocs_emin} \\ 
 &= \frac{E_{(N,\Delta \bar{N})}(\bar{\nu}_0^*,\bar{\nu}_1^*) - \bar{\nu}_0^*N_0 -  \bar{\nu}_1^*N_1  - E_{(N,\Delta N)}(\nu_0^*,\nu_1^*) + \nu_0^*N_0 +  \nu_1^*N_1}{\bar{N}_0-N_0}.\label{jumplocs}
\end{align}
In the second line we have explicitly substituted $p$ and used the observation that Eq. (\ref{Eblock}) allows to determine $E_\mathrm{min}(W_\mathrm{aux}|_{(N,\Delta N)})$ ($E_\mathrm{min}(W_\mathrm{aux}|_{(N,\Delta \bar{N})})$) at arbitrary values $\nu_0^*$, $\nu_1^*$ ($\bar{\nu}_0^*$, $\bar{\nu}_1^*$). The isospin number as well as the ground state energies can be extracted from our simulations, where the former can be determined exactly as the Hamiltonian conserves $N_0$ and $N_1$. Hence the precision of $(\nu_1-\nu_0)|_\mathrm{jump}$ only depends on the precision obtained for the ground state energies. Assuming a systematic error of $\Delta E$ in the energies, one obtains for the error of the location of the phase transition
\begin{align}
 \Delta(\nu_1-\nu_0)|_\mathrm{jump} &= \left|\frac{\partial(\nu_1-\nu_0)|_\mathrm{jump}}{\partial E_{(N,\Delta \bar{N})}(\bar{\nu}_0^*,\bar{\nu}_1^*)}\Delta E_{(N,\Delta \bar{N})}(\bar{\nu}_0^*,\bar{\nu}_1^*)\right| + \left|\frac{\partial(\nu_1-\nu_0)|_\mathrm{jump}}{\partial E_{(N,\Delta N)}(\nu_0^*,\nu_1^*)}\Delta E_{(N,\Delta N)}(\nu_0^*,\nu_1^*)\right| \nonumber \\
 &= \frac{1}{\left|p_{(N,\Delta \bar{N})} - p_{(N,\Delta N)}\right|}\Bigl(\left|\Delta E_{(N,\Delta \bar{N})}(\bar{\nu}_0^*,\bar{\nu}_1^*)\right| +\left| \Delta E_{(N,\Delta N)}(\nu_0^*,\nu_1^*)\right| \Bigr).\label{jumplocs_error}
\end{align}

In practice, we select for each combination of volume and lattice spacing $(Lg,x)$ a single data point inside of each of the phases, where we determine $N_0$ and $N_1$ and estimate the exact energy value as described in the next paragraph. Subsequently, we can compute the location of the phase transition and estimate the error using Eqs. (\ref{jumplocs}) and (\ref{jumplocs_error}).

\section{Extrapolation procedure\label{app:extrapolation}}
\begin{figure}[htp!]
  \centering
  \includegraphics[width=0.5\textwidth]{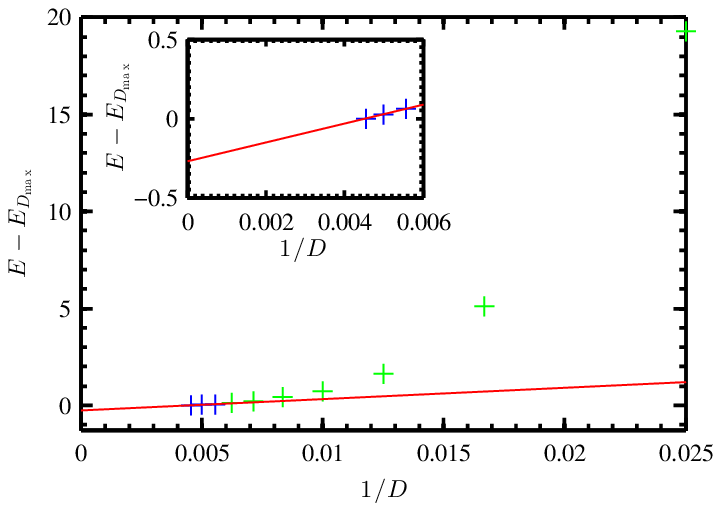}
\caption{Extrapolation in bond dimension for $m/g=0$, $\mu_I/2\pi=0.8$, $x=121$ and $Lg=8$. The blue data points are the ones used for the extrapolation to the limit $D\to\infty$ and the red line shows the linear fit through the blue data points. The inset shows the region close to the origin in better detail.}
\label{fig:extD}
\end{figure}

As explained in the previous paragraph, the precision obtained for the phase transition locations crucially depends on the precision of the ground state energies. To get precise estimates for the exact energy, we extrapolate the bond dimension $D\to\infty$. To do so, we repeat the calculation for each data point for a given combination of volume $Lg$, lattice spacing $x$ and chemical potential difference $\mu_I/2\pi$ for several bond dimensions until the energy approximately scales linearly in $1/D$. For the data presented in the main text, we find that for $x\in[9,36]$ a maximum bond dimension of $D=160$ is enough to enter the linear scaling region, whereas for larger values of $x$ we have to increase the bond dimension up to $220$. Once we enter this regime, we take the last three data points to extrapolate linearly (see Fig.~\ref{fig:extD} for an example). As an estimate for the exact energy we take the mean value of our data point computed with the largest bond dimension, $E_{D_\mathrm{max}}$, and $E_{D=\infty}$ obtained by our extrapolation. The error is estimated as $\Delta E =\frac{1}{2}(E_{D_\mathrm{max}} -E_{D=\infty})$. 

\begin{figure}[htp!]
\centering
\subfigure[Continuum limit for the location of the first phase transition.]{\includegraphics[width=0.48\textwidth]{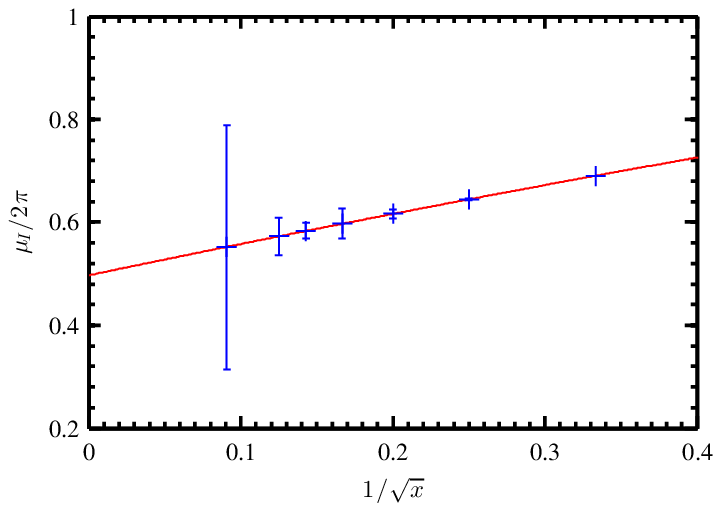}}
\subfigure[Continuum limit for the location of the third phase transition.]{\includegraphics[width=0.48\textwidth]{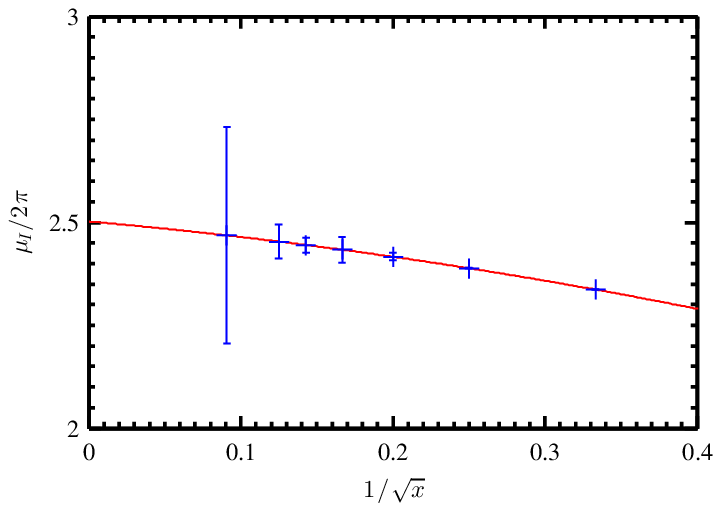}}
\caption{Extrapolation of the phase transition points to the continuum for $Lg=8$ in the massless case. The red line represents a  second order polynomial fit in $1/\sqrt{x}$ and the continuum limit is estimated by taking the value of the fit function at $1/\sqrt{x}=0$.}
\label{fig:jump_continuum_limit}
\end{figure}

In a final step we can now extrapolate the estimated locations for the phase transitions, obtained by the procedure explained in the previous paragraph, to the continuum. We proceed in a standard manner and fit a second order polynomial in $1/\sqrt{x}$ and take the intersection point with the $y$-axis as estimate for the continuum value (see Fig.~\ref{fig:jump_continuum_limit} for an example). As an error estimate for the continuum value, we take the fitting error where we use a $1\sigma$ confidence interval.
The final results for the location of the phase transitions obtained after the full extrapolation procedure are shown in Tabs. \ref{tab:res_mg0} - \ref{tab:res_mg05}.
\begin{table}[H]
\centering
 \begin{tabular}{|c||c|c|c|c|}
 \hline
 Volume & 1. transition & 2. transition & 3. transition & 4. transition \\ \hline
 $Lg=2$  & $0.499960(88)$ & $1.513345(47)$ & $2.617208(11)$ & $3.716041(12)$ \\ \hline
 $Lg=6$  & $0.499(21)$ & $1.501(23)$ & $2.504(22)$ & $3.511(20)$ \\ \hline
 $Lg=8$  & $0.497(49)$ & $1.501(60)$ & $2.502(55)$ & $3.505(51)$ \\ \hline
 \end{tabular}
\caption{Continuum estimates for the locations of the first four phase transitions for the massless case $m/g=0$.}
\label{tab:res_mg0}
\end{table}

\begin{table}[H]
\centering
 \begin{tabular}{|c||c|c|c|c|}
 \hline
 Volume & 1. transition & 2. transition & 3. transition & 4. transition \\ \hline
 $Lg=2$  & $0.522620(86)$ & $1.515910(40)$ & $2.620237(14)$ & $3.716558(20)$ \\ \hline
 $Lg=6$  & $0.711(19)$ & $1.538(26)$ & $2.519(23)$ & $3.520(20)$ \\ \hline
 $Lg=8$  & $0.831(42)$ & $1.575(65)$ & $2.532(57)$ & $3.523(52)$ \\ \hline
 \end{tabular}
\caption{Continuum estimates for the locations of the first four phase transitions for $m/g=0.125$.}
\label{tab:res_mg0125}
\end{table}

\begin{table}[H]
\centering
 \begin{tabular}{|c||c|c|c|c|}
 \hline
 Volume & 1. transition & 2. transition & 3. transition & 4. transition \\ \hline
 $Lg=2$  & $0.554897(76)$ & $1.522594(40)$ & $2.624794(14)$ & $3.720370(19)$ \\ \hline
 $Lg=6$  & $0.938(16)$ & $1.617(26)$ & $2.558(23)$ & $3.546(20)$ \\ \hline
 $Lg=8$  & $1.165(39)$ & $1.728(66)$ & $2.606(57)$ & $3.571(52)$ \\ \hline
 \end{tabular}
\caption{Continuum estimates for the locations of the first four phase transitions for $m/g=0.25$.}
\label{tab:res_mg025}
\end{table}

\begin{table}[H]
\centering
 \begin{tabular}{|c||c|c|c|c|}
 \hline
 Volume & 1. transition & 2. transition & 3. transition & 4. transition \\ \hline
 $Lg=2$  & $0.643234(66)$ & $1.548542(35)$ & $2.644094(11)$ & $3.732926(20)$ \\ \hline
 $Lg=6$  & $1.402(12)$ & $1.874(23)$ & $2.703(22)$ & $3.647(20)$ \\ \hline
 $Lg=8$  & $1.816(24)$ & $2.168(53)$ & $2.871(55)$ & $3.752(49)$ \\ \hline
 \end{tabular}
\caption{Continuum estimates for the locations of the first four phase transitions for $m/g=0.5$.}
\label{tab:res_mg05}
\end{table}

\section{Effect of non-vanishing mass on the phase structure}
\begin{figure}[htp!]
\centering
\subfigure[Volume $Lg=2$.]{\includegraphics[width=0.48\textwidth]{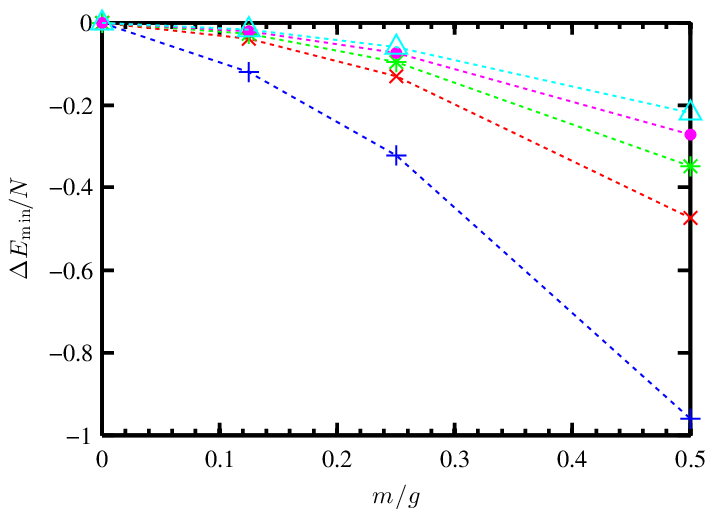}}
\subfigure[Volume $Lg=6$.]{\includegraphics[width=0.48\textwidth]{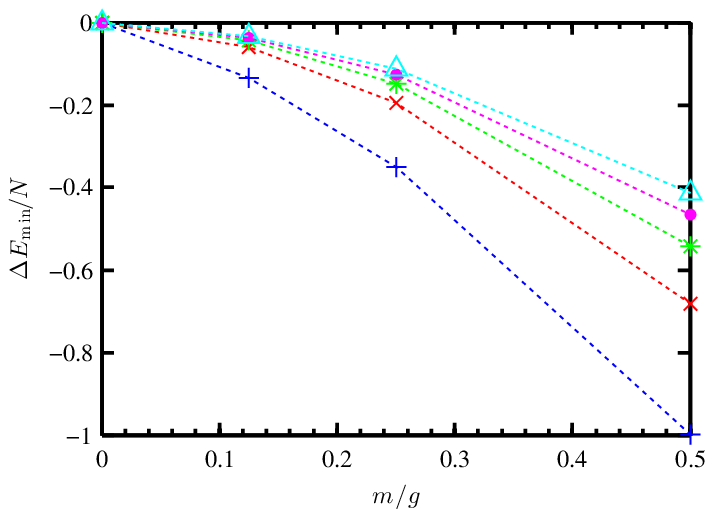}}
\subfigure[Volume $Lg=8$.]{\includegraphics[width=0.48\textwidth]{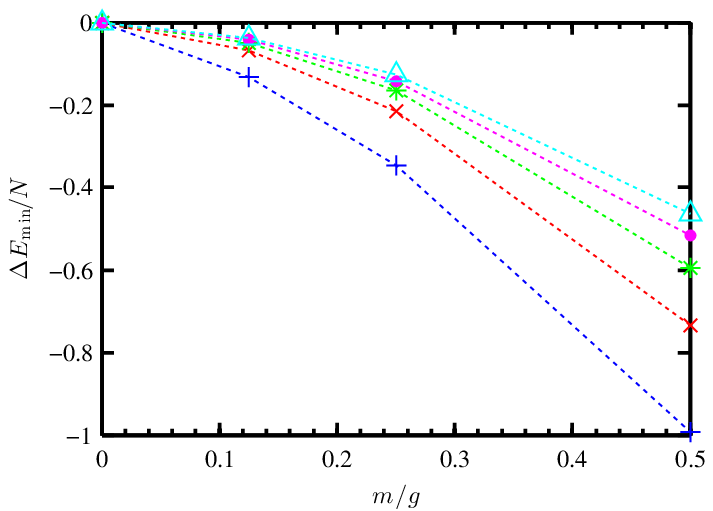}}
\caption{Energy shift per site $\Delta E_\mathrm{min}/N$ as a function of $m/g$ for $x=121$ and volumes $Lg=2$ (a), $Lg=6$ (b) and $Lg=8$ (c). The different markers indicate the different phases characterized by the isospin number, blue crosses represent $\Delta N=0$, red \X's $\Delta N=2$, green asterisks $\Delta N=4$, magenta dots $\Delta N=6$ and cyan triangles $\Delta N=8$. As a guide for the eye the data points are connected with dotted lines.}
\label{fig:dE_over_N}
\end{figure}

Figures \ref{fig:jumplocsBond_mg05} and \ref{fig:phasediagram} in the main text, as well as the Tabs.~\ref{tab:res_mg0125} - \ref{tab:res_mg05}, show that for non-vanishing fermion mass the locations of the phase transitions between phases characterized by small $\Delta N$ are affected the most compared to the massless case. Transitions between phases with larger isospin number are less influenced and only slightly shifted towards higher values of $\mu_I/2\pi$. This behavior can be explained qualitatively with a change in $E_\mathrm{min}(W_\mathrm{aux}|_{(N,\Delta N)})$ which is the only mass dependent contribution to the energy, as can be seen from Eqs. (\ref{Eblock}) and (\ref{Eblockscaling}). Consequently introducing a nonzero value for $m/g$ leads to a shift $\Delta E_\mathrm{min}$ with respect to the massless case, $E_\mathrm{min}(W_\mathrm{aux}|_{(N,\Delta N)}) = E_\mathrm{min}(W_\mathrm{aux}|_{(N,\Delta N)})|_{m/g=0} +\Delta E_\mathrm{min}$. Equation (\ref{jumplocs_emin}) reveals that these energy shifts affect the locations of the phase transitions, as soon as they are not equal in every phase. Extracting $\Delta E_\mathrm{min}/N$ inside each phase for our smallest lattice spacing for several masses, we obtain the results shown in Fig. \ref{fig:dE_over_N}, which clearly show that the shifts are different for each phase. In particular, we see that for the phase characterized by $\Delta N=0$, the energy shift is a lot more pronounced than for the phase characterized by $\Delta N=2$, thus explaining the significant shift towards higher values of $\mu_I/2\pi$ for the location of the first phase transition with respect to the massless case. For phases with larger isospin number, the energy shifts differ less, consistent with the observation that the locations of the phase transitions between these phases are less affected. Although for all three volumes studied we observe similar energy shifts, Fig. \ref{fig:jumplocsBond_mg05} as well as Tabs.~\ref{tab:res_mg0125} - \ref{tab:res_mg05} show that for $Lg=2$ the locations of the phase transitions are less affected by a nonzero fermion mass. This is likely due to the finite volume effects arising from the fact that the total fermion number corresponds to the number of sites as described in the main text.

\end{document}